\documentclass[useAMS,usenatbib]{mn2e}

\usepackage{psfig}
\usepackage{amsmath}
\usepackage{amssymb}
\usepackage{upgreek}
\usepackage{longtable}
\usepackage[mathscr]{eucal}
\usepackage[dvips]{graphicx}

\title
[Asymmetric expansion of the Lagoon Nebula cluster]
{The {\it Gaia}-ESO Survey: Asymmetric expansion of the Lagoon Nebula cluster NGC 6530 from GES and {\it Gaia} DR2}
\author
[Wright et al.]
{Nicholas J. Wright$^1$, R.D. Jeffries$^1$, R.J. Jackson$^1$, A. Bayo$^{2,3}$, R. Bonito$^4$, \newauthor
F. Damiani$^4$, V. Kalari$^5$, A.C. Lanzafame$^6$, E. Pancino$^{7,8}$, R.J. Parker$^{9,}$\thanks{Royal Society Dorothy Hodgkin Fellow}, \newauthor
L. Prisinzano$^4$, S. Randich$^7$, J.S. Vink$^{10}$, E.J. Alfaro$^{11}$, M. Bergemann$^{12}$, \newauthor
E. Franciosini$^7$, G. Gilmore$^{13}$, A. Gonneau$^{13}$, A. Hourihane$^{13}$, P. Jofr\'e$^{14}$, \newauthor
S.E. Koposov$^{13}$, J. Lewis$^{13}$, L. Magrini$^7$, G. Micela$^4$, L. Morbidelli$^7$, \newauthor
G.G. Sacco$^7$, C.C. Worley$^{13}$ and S. Zaggia$^{15}$. \\
\\
$^{1}$Astrophysics Group, Keele University, Keele, ST5 5BG, UK\\
$^{2}$Instituto de F\'isica y Astronomi\'ia, Universidad de Valpara\'iso, Chile\\
$^{3}$N\'ucleo Milenio Formaci\'on Planetaria - NPF, Universidad de Valpara\'iso, Av. Gran Breta\~na 1111, Valpara\'iso, Chile\\
$^{4}$INAF - Osservatorio Astronomico di Palermo, Piazza del Parlamento 1, 90134, Palermo, Italy\\
$^{5}$Departamento de Astronom\'ia, Universidad de Chile, Casilla 36-D, Santiago, Chile\\
$^{6}$Dipartimento di Fisica e Astronomia, Sezione Astrofisica, Universit\'{a} di Catania, via S. Sofia 78, 95123, Catania, Italy\\
$^{7}$INAF - Osservatorio Astrofisico di Arcetri, Largo E. Fermi 5, 50125, Florence, Italy\\
$^{8}$Space Science Data Center - Agenzia Spaziale Italiana, via del Politecnico, s.n.c., I-00133, Roma, Italy\\
$^{9}$Department of Physics \& Astronomy, University of Sheffield, Sheffield, S3 7RH, UK\\
$^{10}$Armagh Observatory, College Hill, Armagh, BT61 9DG, Northern Ireland, UK\\
$^{11}$Instituto de Astrof\'isica de Andaluc\'ia (CSIC), Glorieta de la Astronom\'ia s/n, Granada 18008, Spain\\
$^{12}$Max-Planck Institut f\"{u}r Astronomie, K\"{o}nigstuhl 17, 69117 Heidelberg, Germany\\
$^{13}$Institute of Astronomy, University of Cambridge, Madingley Road, Cambridge CB3 0HA, United Kingdom\\
$^{14}$N\'ucleo de Astronom\'ia, Universidad Diego Portales, Av. Ejercito 441, Santiago, Chile\\
$^{15}$INAF - Osservatorio Astronomico di Padova, Vicolo della Osservatorio 5, 35122 Padova, Italy
}

\begin{document}
\maketitle

\begin{abstract}

The combination of precise radial velocities from multi-object spectroscopy and highly accurate proper motions from {\it Gaia} DR2 opens up the possibility for detailed 3D kinematic studies of young star forming regions and clusters. Here, we perform such an analysis by combining {\it Gaia}-ESO Survey spectroscopy with {\it Gaia} astrometry for $\sim$900 members of the Lagoon Nebula cluster, NGC~6530. We measure the 3D velocity dispersion of the region to be $5.35^{+0.39}_{-0.34}$~km~s$^{-1}$, which is large enough to suggest the region is gravitationally unbound. The velocity ellipsoid is anisotropic, implying that the region is not sufficiently dynamically evolved to achieve isotropy, though the central part of NGC~6530 does exhibit velocity isotropy that suggests sufficient mixing has occurred in this denser part. We find strong evidence that the stellar population is expanding, though this is preferentially occurring in the declination direction and there is very little evidence for expansion in the right ascension direction. This argues against a simple radial expansion pattern, as predicted by models of residual gas expulsion. We discuss these findings in the context of cluster formation, evolution and disruption theories.

\end{abstract}


\section{Introduction}

Stars form within turbulent and clumpy giant molecular clouds that result in a hierarchical and highly-substructured spatial distribution of the youngest stars and protostars \citep[e.g.,][]{elme02,gute08}. The spatial distribution of very young stars, still in their natal star forming regions, often retains this substructure \citep{lars95,cart04,gute08}, though many centrally-concentrated and smoothly-distributed star clusters are also observed at this early time \citep[e.g.,][]{carp00,pfal09}. Once young stars have spatially decoupled from their birth environment the majority are found in unbound groups of some sort \citep{lada03}, often in the form of associations \citep[e.g.,][]{brow97,wrig16}.

The physical processes responsible for this evolution and the progression of stars from their formation sites to the Galactic field are not well constrained as there are many different processes that can play a role. The initial spatial and kinematic structure in which stars form can play a significant part in their future evolution \citep{park14,park16}, as can dynamical interactions between stars and binaries \citep[e.g.,][]{mark12,sill18}, and the evolving gravitational potentials of the stellar and gaseous parts of the system \citep[e.g.,][]{baum07,moec10}. The latter issue has been argued to be critical to the survival or dispersal of young star clusters following residual gas expulsion \citep[e.g.,][]{tutu78,lada84}, though observational kinematic evidence has yet to verify this picture \citep{wrig16,wrig18} and there are questions over the efficiency and effectiveness of this process \citep{krui12b,giri12,dale15}. The kinematics of young stars can provide important tests of these theories, allowing us to follow the motions of stars as stellar systems form, interact and dissolve.

We are currently undergoing a transformative improvement in kinematic data quantity and quality. This is driven by data from current radial velocity (RV) surveys \citep[such as the 340-night {\it Gaia}-ESO Survey, GES,][]{gilm12,rand13} and astrometric data of unprecedented precision from {\it Gaia} \citep{prus16}. Individually these datasets have provided insights into the structure and dynamics of young star forming regions \citep{jeff14,rigl16,berl19}, the kinematics of runaway OB stars \citep{drew18}, and revealed the low-mass stellar content of dispersed OB associations and star-forming complexes \citep{zari17,zari18,arms18,becc18}. However, the real value comes from combining these data to achieve 3D kinematics, providing estimates of energies, angular momenta and dynamics, and avoiding the need to make crude isotropy assumptions. In this paper we combine GES RVs and {\it Gaia} proper motions (PMs) for hundreds of young stars in the Lagoon Nebula, using both spectroscopic and X-ray diagnostics of youth to compile a kinematically-unbiased sample of members.

\begin{figure*}
\begin{center}
\includegraphics[width=500pt]{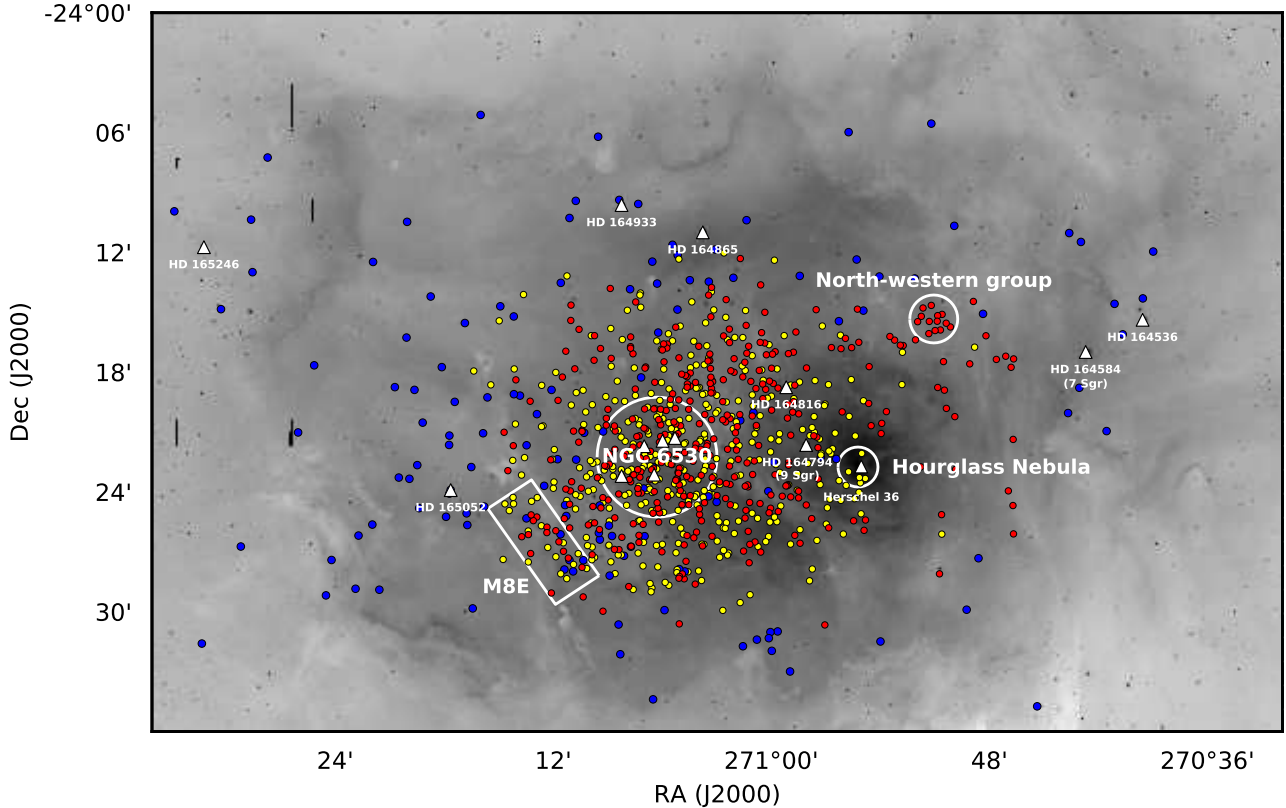}
\caption{Map of the spatial distribution of our final sample of young stellar objects projected onto an inverted VPHAS+ H$\alpha$ image. Sources are colour-coded by their origin: blue sources are selected based only on their spectroscopic GES information, red sources only on their MYSTIX membership probabilities, and yellow sources selected using both GES and MYSTIX information. The central NGC~6530 cluster, the Hourglass Nebula, the eastern `M8E' group and the north-western group of stars are indicated based on locations from \citet{toth08} and \citet{kuhn14}. The high-mass stars discussed in Section~\ref{s-highmass} are indicated with white triangles and labelled (HD 164906, HD 315031, CD-24 13829, CPD-24 6164, and CD-24 13837 fall within the NGC 6530 circle and are not labelled for clarity).}
\label{map_lagoon}
\end{center}
\end{figure*}

NGC~6530 is a young \citep[1--2~Myr,][]{mayn07,bell13} cluster projected against the centre of the Lagoon Nebula H{\sc ii} region (Messier 8), the illuminated part of a giant molecular cloud that extends several degrees \citep{lada76,sung00}, or $\sim$50~pc at the estimated distance of 1326~pc (see Section~\ref{s-combine}). The main sources of ionization are the massive stars 9~Sgr (HD~164794, an O4V(f) star) and the massive binary HD~165052 (O6.5V + O7.5V), as well as a number of late O and B-type stars \citep[see e.g.,][and Figure~\ref{map_lagoon}]{toth08}. A rich population of relatively unreddened low-mass members has also been uncovered using optical and X-ray data \citep{sung00,dami04,pris05,kala15}. The youngest stars in the region are associated with embedded populations in the Hourglass Nebula, which houses an ultra-compact H{\sc ii} region, and the M8~E region around the massive protostar M8E-IR \citep{toth08}. The spatial arrangement of these separate populations is shown in Figure~\ref{map_lagoon}. A recent review of the Lagoon Nebula can be found in \citet{toth08}.

In this paper we present the first 3D kinematic study of the Lagoon Nebula stellar population and its central cluster NGC~6530. In Section~2 we outline the observational data used and in Section~3 we discuss how a kinematically unbiased sample of members was compiled for our study. In Section~4 we present our results, including the calculation of the velocity dispersions, the search for velocity gradients, the assessment of the evidence for expansion and rotation, and a consideration of the kinematic outliers in the sample. In Section~5 we discuss our results and their implications for the understanding of young star clusters and in Section~6 we summarise our results.

\section{Observational data}

In this Section we describe the observational data used in this study, including GES spectroscopy, {\it Gaia} data release 2 (DR2) astrometry, and ancillary membership information from X-ray, infrared and H$\alpha$ surveys.

\subsection{{\it Gaia}-ESO Survey spectroscopy}

The main source of data for this study is spectroscopy from GES observations of NGC~6530. The GES target selection strategy for clusters is carried out independently, but following homogeneous criteria, for each cluster observed during the survey. The targets for these observations were selected to cover an area of $55^\prime \times 30^\prime$ in the direction of NGC~6530, centred on $(\alpha, \delta) = (271.09, -24.33)$. The targets were chosen based on their location in the $V$ vs $V-I$ colour-magnitude diagram (CMD) using photometry from \citet{pris05} and the $r^\prime$ vs $r^\prime - i^\prime$ CMD from \citet{drew14}. Targets were selected that fell between the 0.5 and 10~Myr \citet{sies00} isochrones at a distance of 1.3~kpc and an extinction of E(B-V) = 0.33, and between $V$ magnitudes of 12 and 19 (corresponding to pre-main sequence masses of 2.5 to 0.4~M$_\odot$ at an age of 2~Myr). This led to a selection of 4066 target candidates that covered a wide area around the expected cluster pre-main sequence in the CMD (see Figure~\ref{cmd}), ensuring that target selection was not biased by the assumed cluster properties.

The targets were grouped according to their $V$ magnitude and configured for multi-fibre spectroscopy across 27 observing blocks. These blocks were observed with the FLAMES\footnote{Fibre Large Array Multi Element Spectrograph} fibre-fed spectrograph at the VLT\footnote{Very Large Telescope} over 17 nights in September 2012 and June to September 2013, resulting in 1872 targets being observed. The targets that were observed are not preferentially brighter or fainter than those that weren't observed, but were just those selected by the fibre allocation software. The only bias that this might lead to is the preference to select stars in non-crowded regions where it can be difficult to position multiple fibres close together. However, thanks to the large number of observing blocks observed this is unlikely to be a major bias. The observations were performed using the GIRAFFE intermediate-resolution spectrograph and the HR15N setup, which gave spectra with a resolving power of 14,000--15,000 covering a wavelength range of 6444--6816~\AA\, and the UVES high-resolution spectrograph with the U520 and U580 setups, which give a resolution of 47,000 and wavelength ranges of 4180--6212 and 4786--6830~\AA\ respectively. See \citet{pris19} for more details.

GES data were reduced and analysed using common methodologies and software to produce a uniform set of spectra and stellar parameters, and are described in \citet{jeff14} and \citet{sacc14} for the data reduction and \citet{lanz15} and \citet{panc17} for the data analysis, calibration and extraction of stellar parameters of stars in young clusters. In this paper we use data from the fifth internal data release (iDR5) of November 2015. In bright H{\sc ii} regions such as the Lagoon Nebula some stellar spectra can be contaminated with nebular emission lines that can affect the calculation of RVs. We therefore compared the RVs calculated from the GES pipeline \citep{jeff14} with those calculated from spectra with nebular lines masked (see Klutsch et al. in prep.) and retained those where the two methods agreed to within 3$\sigma$ (75\% of GES RVs pass this test).

We also include in our analysis VLT/FLAMES observations of 228 stars in NGC~6530 contained in the ESO Archive. These observations, performed on May 27, 2003, were originally presented by \citet{pris07}, with targets selected from photometry presented by \citet{pris05} and falling in the $V$ vs $V-I$ CMD in the vicinity of X-ray selected members of the cluster \citep{dami04} and with $V$ magnitudes between 14.0 and 18.2~mag. These data were re-reduced and analysed as part of GES following the same methodologies described above.

Combining GES and archival data and removing duplicates we have spectroscopy for 1957 objects. Out of these objects we exploit measured effective temperature for 1203 stars (61\% of the sample), lithium equivalent widths for 1545 stars (79\%), and RVs for 1326 stars (68\%).

\subsection{{\it Gaia} DR2 astrometry}
\label{s-gaia_dr2}

We use astrometric data from {\it Gaia} \citep{prus16} DR2 \citep{brow18}, which contains parallaxes, $\varpi$, and PMs, $\mu_\alpha$\footnote{We follow standard practice and provide the PM in right ascension as $\mu_\alpha = \mu_{\alpha,0} \, \mathrm{cos} \, \delta$.} and $\mu_\delta$, calculated from the first 22 months of {\it Gaia} observations (2014--2016). This data achieves a very high level of astrometric precision for a sample of unprecedented size, though it is still calculated assuming single-star behaviour \citep{lind18}, meaning that close binaries (separation $\leq$ 100 mas) will be unresolved and the astrometry is for their photocentre. The astrometric parameters are aligned with the International Celestial Reference Frame to a precision of about 0.02~mas at epoch J2015.5 \citep[see][for details]{lind18} and are absolute \citep[they do not rely on an external reference frame and no significant residual offsets exist,][]{aren18}.

DR2 is effectively complete in the range $G = 12 -- 17$~mag, with a detection threshold of $G = 20.7$~mag. Very bright ($G < 7$~mag) and high proper-motion ($> 0.6$~arcsec/yr) stars suffer from incompleteness, though this should not affect us. In dense areas of the sky ($> 400,000$~stars~deg$^{-2}$) the effective magnitude limit of the survey can drop to $\sim$18~mag and chance configurations can lead to confusion and mistakenly large parallaxes \citep{lind18}, though the stellar density of 220,000~stars~deg$^{-2}$ towards NGC~6530 is not high enough to make this an issue. Spatial correlations in parallax and PM are believed to exist on small ($<1$~deg) and intermediate ($\sim$20~deg) angular scales up to $\pm$0.1~mas for parallaxes and $\pm$0.1~mas~yr$^{-1}$ for PMs \citep{lind18,luri18}. As a result of this, averaging parallaxes over small regions of the sky will not reduce the parallax uncertainty on the mean to below this \citep{luri18}. These systematic uncertainties can be estimated using tables of spatial covariances\footnote{Available from https://www.cosmos.esa.int/web/gaia/dr2-known-issues.}. Comparison with known quasars suggests the parallaxes have an overall negative bias of $0.03 \pm 0.02$~mas \citep[i.e., 0.03~mas should be added to published values when calculating distances,][]{lind18} and that the uncertainties may be under-estimated by 8-12\%, or higher for bright stars and regions of higher density \citep{aren18}. We do not use stellar effective temperatures from DR2 because they are derived under the assumption of no reddening \citep{andr18,aren18}, which is likely to be invalid for our targets. 

{\it Gaia} DR2 data was downloaded in a circular region of radius 0.6~deg centred on NGC~6530, covering the existing GES observations, and including 249,351 sources. We filtered this source list based on two criteria outlined in the {\it Gaia} data release papers and technical notes, as described below, but did not apply other cuts based on absolute astrometric uncertainties or negative parallaxes, both of which have been shown to introduce significant biases \citep{aren18}. First we required that all sources have more than 8 visibility periods used for their astrometric solution, which removes sources with spurious PMs and parallaxes \citep{aren18}. Then we applied the cut recommended in \citet{lind18b} using the `re-normalised unit weighted error' (RUWE), requiring that $\mathrm{RUWE} \leq 1.4$, and using the normalisation factors provided by {\it Gaia}'s Data Processing and Analysis Consortium (DPAC). This removes sources with spurious astrometry and helps filter contamination from double stars, astrometric effects from binary stars and other contamination problems, and replaces all previously-recommended astrometric filters \citep[e.g., those outlined in][]{lind18}. We did not implement the photometric cut in equation C-2, since photometry is not the main use we will make of the {\it Gaia} data. Together the cuts remove 69\% of {\it Gaia} sources, the majority of which are faint, leaving 76,554 sources.

We experimented with an additional cut using the astrometric excess noise, which is a measure of the astrometric goodness of fit calculated as the extra noise per observation that has been included in the astrometric uncertainties to explain the scatter of residuals in the astrometric solution \citep{lind18}.  An outright cut using this parameter has not been recommended when using {\it Gaia} DR2 data since this quantity is less discriminating than other cuts due to a bug in DR2 and because cuts based purely on astrometric uncertainties can lead to a biased sample \citep{luri18}. Despite this we repeated all astrometric analyses in this paper using a subset of our final sample made up only of sources with astrometric excess noise equal to zero. We found that our results and conclusions did not change and only that the uncertainty on many of our measured quantities actually increased, because of the significantly smaller sample used for this analysis.

The primary application of these data will be to use PMs to study the plane-of-the-sky kinematics alongside the line-of-sight kinematics from the RVs. We will briefly use the parallaxes to refine our membership and identify foreground and background contaminants, but will not exploit these data further. In the astrometric analysis that follows we always use the full ($5 \times 5$) {\it Gaia} covariance matrix whenever propagating uncertainties, and consider all errors to follow a normal distribution \citep{aren18}. In addition we increase the astrometric errors by factors of 1.093 ($\mu_\alpha$), 1.115 ($\mu_\delta$), and 1.081 ($\varpi$), in accordance with the extent to which the formal uncertainties appear to have been underestimated \citep{lind18}.

\subsection{Supporting data}

\begin{figure}
\begin{center}
\includegraphics[height=240pt, angle=270]{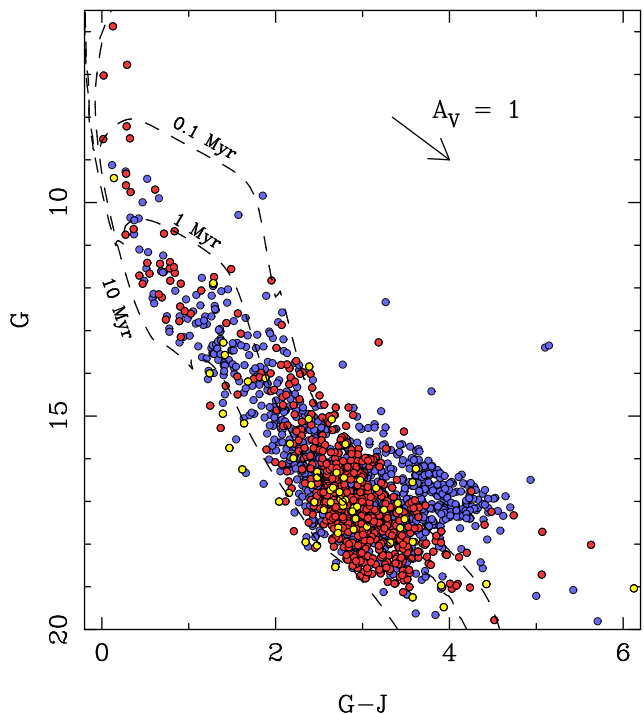}
\caption{$G$ vs $G-J$ colour magnitude diagram showing 1914 candidate members of NGC~6530 (those with both {\it Gaia} and 2MASS photometry from the 4066 target candidates considered for GES observations) in blue and over-plotted in red and yellow are 683 of the 889 stars from our final list of members (those with both {\it Gaia} and 2MASS photometry). Red points are members of the Lagoon Nebula population in the main velocity distribution, while yellow points are PM kinematic outliers (see Section~\ref{s-kinematicoutliers}). Also shown are a 1~mag extinction vector and pre-main sequence stellar isochrones from \citet{mari17}, the latter reddened by $A_V = 1.1$~mag \citep{sung00} and at a distance of 1326~pc (see Section~\ref{s-combine}).}
\label{cmd}
\end{center}
\end{figure}

We complement the spectroscopic and astrometric data with supporting literature data to aid the identification of young stars across the Lagoon Nebula. This includes membership probabilities for 2427 sources from the Massive Young star-forming Complex Study in Infrared and X-rays \citep[MYStIX,][]{feig13} presented by \citet{broo13}, and VST Photometric H$\alpha$ Survey \citep[VPHAS,][]{drew14} photometric H$\alpha$ equivalent widths for 235 sources from \citet{kala15}. Details of how these data were incorporated into our membership selection can be found in the next section. We also use photometry from the 2 Micron All Sky Survey \citep[2MASS,][]{cutr03}. All catalogues are cross-matched with each other using a 1$^{\prime\prime}$ matching radius. A subset of our candidate source list (those with both {\it Gaia} and 2MASS photometry) is shown in Figure~\ref{cmd}.

\section{Membership selection}

In this section we outline the various indicators of stellar youth used to identify young stars in the direction of NGC~6530. To avoid any kinematic biases we have not used any sort of membership cut based on RVs or PMs.

\subsection{Spectroscopic membership indicators}
\label{s-spec}

GES spectroscopy facilitates three separate indicators of stellar youth: the gravity index $\gamma$, the EW(Li), and the full width at zero intensity (FWZI) of the H$\alpha$ line. These three parameters have been used by multiple GES studies to define samples of young stars free of kinematic bias \citep[e.g.,][]{jeff14,rigl16,sacc17,brav18}, and we follow their methods here.

\begin{figure*}
\begin{center}
\includegraphics[height=500pt, angle=270]{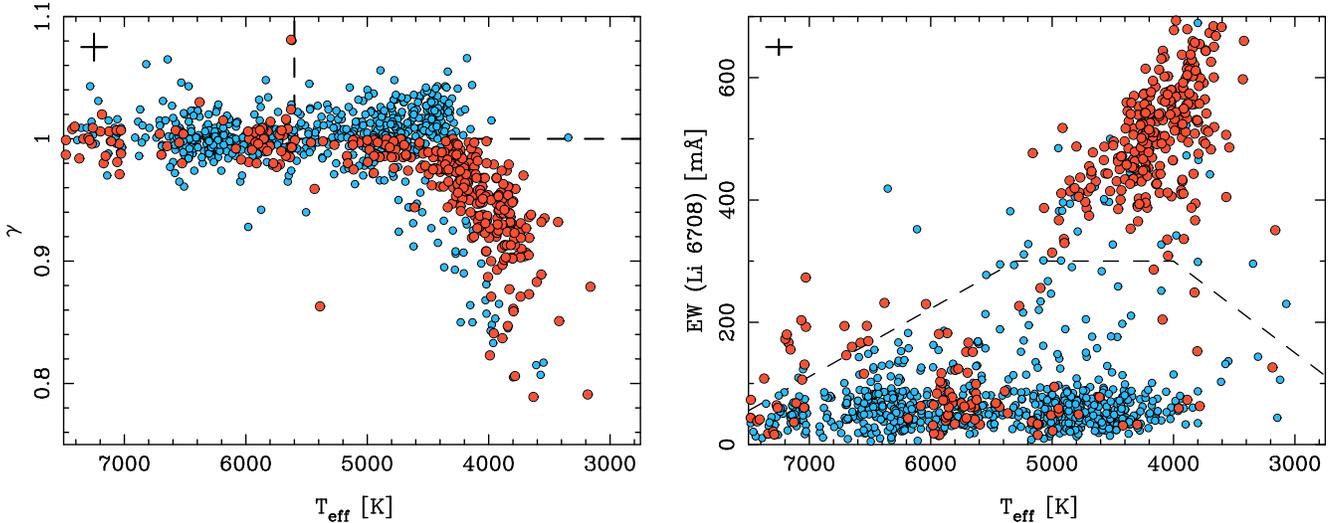}
\caption{GES spectral quantities used for membership selection. The gravity index $\gamma$ (left panel) and the EW of the lithium 6708~\AA\ line (right panel) are shown as a function of effective temperature. In both panels red circles show sources from our final list of members and blue circles show non-members, as determined using both of these diagnostics and H$\alpha$ information. The dashed lines show the thresholds used to identify giants in the upper-right corner of the $\gamma$-$T_{eff}$ plot and to identify young stars above the threshold in the EW(Li)-$T_{eff}$ plot (stars can fall below this line and still be considered members if they exhibit H$\alpha$ emission). A typical error bar is shown in the top-left corner of each panel illustrating the typical uncertainties of 100~K in $T_{eff}$, 13~m\AA\ for EW(Li) and 0.011 for $\gamma$.}
\label{GES_spectral}
\end{center}
\end{figure*}

One of the main sources of contamination in a sample of candidate members selected from the CMD is background giants. These can be identified and removed based on their low surface gravity using the gravity index $\gamma$ \citep{dami14}\footnote{While pre-main sequence stars do have lower gravity than main sequence stars it is not as low as for background giants and this method has successfully been used to separate young stars in Chamaeleon~I from background stars \citep{sacc17}.}. Figure~\ref{GES_spectral} shows $\gamma$ as a function of the effective temperature for GES targets observed towards NGC~6530. Giant stars occupy the upper-right part of this plot, with $\gamma > 1$ and $T_{eff} < 5600$~K \citep{dami14}. There are 1064 stars with $\gamma$ and $T_{eff}$ measurements in our sample, 233 of which meet these criteria and are classified as giants. A further 13 stars lack $T_{eff}$ measurements and of these, 5 have $\gamma > 1$ and are also classified as giants. This leaves 839 candidate young stars.

The other major source of contamination is main-sequence stars in the foreground of NGC~6530. These sources can be separated from the young stars in NGC~6530 based on the presence of lithium in their atmospheres, since late-type stars deplete their lithium after approximately 10 (for M-type stars) to 100~Myr (for K-type stars) due to burning and subsequent mixing throughout the convection zone \citep{sode10}. There are 827 non-giant stars with EW(Li) and $T_{eff}$ measurements in our sample (field giants can show lithium enrichment and so both membership criteria are required), as shown in Figure~\ref{GES_spectral}. Of these, 358 have EW(Li) above the threshold defined by \citet{sacc17} based on the distribution of EW(Li) values for stars in the 30--50~Myr cluster IC~2602 \citep{rand97}. A further 8 candidate young stars lack $T_{eff}$ measurements, but none of these have EW(Li) above the highest threshold value of 300~m\AA\ and so are not considered members.

The EW(Li) can be underestimated in stars with high mass accretion rates if the continuum emission in excess produced by the accretion shock reduces the measured signal \citep[e.g.,][]{pall05}. From the 481 stars that fail the EW(Li) membership test, 82 have FWZI(H$\alpha$) measures greater than 4~\AA\ \citep{boni13,pris19} and 3 have photometric H$\alpha$ equivalent widths above the threshold defined by \citet{barr03}, and are therefore re-classified as members. This leaves 443 stars as high-confidence young stars based on their spectroscopic properties.

\subsection{Non-spectroscopic membership indicators}
\label{s-nonspec}

\citet{broo13} estimate membership probabilities for stars towards NGC~6530 based on a combination of X-ray, near-IR, and mid-IR data. They employ a ``Naive Bayes Classifier'' that estimates probabilities that each source is either a young star in the star forming region, a foreground or background field star, or an extragalactic object. These probabilities are estimated based on key observational quantities and likelihood models for each quantity and population. An object is assigned a membership class if the probability of the object being a member of that population is more than twice the probability of it being in any other population. The value of this dataset is not just the membership information, but also the published probabilities for each source that it is a member of a given population, which we can combine with other (e.g., spectroscopic) information to produce an improved membership probability (Section~\ref{s-combine}).

\citet{broo13} include 2427 sources in the direction of NGC~6530, of which 1828 are classified as young stars, 2 and 3 are foreground and background stars, respectively, 102 are extragalactic sources, and 492 are unclassified. Sources may be unclassified either because no individual population class has a probability more than twice as large as the next-largest probability, or if a source was initially classified as a member but lacks observational evidence for being a member other than the position-dependent prior. The total number of sources not classified as members (including unclassified sources) is consistent with the predicted number of contaminants from their simulations, which they argue supports their classification.

\subsection{Known high-mass members}
\label{s-highmass}

Previously-known high-mass members of NGC~6530 and the surrounding Lagoon Nebula were gathered from the literature \citep[e.g.,][and others]{toth08} and are listed in Table~\ref{high_mass_members}. Spectral types, RVs and photometry were gathered from the literature, while parallaxes, PMs and astrometric parameters were taken from {\it Gaia} DR2. Five of these stars were found to have poor DR2 astrometry (three lacking sufficient observing periods and two having high values of RUWE, see Section~\ref{s-gaia_dr2}), and since these sources also lack literature RVs they were not included in our kinematic analysis. The star 7~Sgr (HD~164584), suspected to be a member of NGC~6530 \citep[e.g.,][]{dami17}, has a large parallax that suggests it is a foreground star, and since it is not thought to be a binary \citep{eggl08} that might have uncertain astrometry, we do not include it in our analysis. The eight remaining OB stars were included in our sample.

\begin{table*}
\caption{Previously known high mass members of the Lagoon Nebula. 
References are GOSS \citep[Galactic O-Star Spectroscopic Survey,][]{sota14}, B89 \citep{bogg89}, G77 \citep{garr77}, G89 \citep{gray89b}, G14 \citep{gonz14}, H65 \citep{hilt65}, H88 \citep{houk88}, and L06 \citep{leve06}. RVs taken from \citet{cont77}, \citet{pour04} and {\it Gaia} DR2. Parallaxes are from {\it Gaia} DR2.}
\label{high_mass_members} 
\begin{tabular}{lllccl}
\hline
Name(s)			& Spectral			& Ref.	& RV				& Parallax				& Included\\
				& type			&		& [km/s]			& [mas]				& in sample?\\
\hline
HD 164794 (9 Sgr)	& O4V			& GOSS	& $-0.8 \pm 1.9$	& $0.851 \pm 0.095$		& Yes \\ 
HD 165052		& O5.5V + O8V		& GOSS	& $1.05 \pm 0.31$	& $0.784 \pm 0.046$		& Yes \\ 
HD 164536		& O7V + O7.5V		& GOSS	& 				& $0.792 \pm 0.251$		& No, poor astrometry \\ 
Herschel 36		& O7.5V			& GOSS	& 				& $0.902 \pm 0.219$		& No, poor astrometry\\ 
HD 165246		& O8V			& GOSS	& 				& $0.501 \pm 0.113$		& No, poor astrometry\\ 
HD 164816		& O9.5V + B0V		& GOSS	& 				& $0.844 \pm 0.070$		& Yes \\ 
HD 164906		& B0Ve			& L06	& 				& $0.810 \pm 0.055$		& No, poor astrometry\\  
HD 164865		& B9Iab			& G77	& 				& $0.715 \pm 0.052$		& Yes\\ 
HD 164584 (7 Sgr)	& F2/3 II/III		& G89	& $-5.62 \pm 0.57$	& $2.805 \pm 0.152$		& No, large parallax\\ 
HD 315031		& B0.5V + B1V		& G14	& 				& $0.690 \pm 0.070$		& Yes\\ 
CD-24 13829		& B1.5V			& H65	&				& $1.000 \pm 0.098$		& No, poor astrometry\\ 
CPD-24 6164		& B1V			& H65	& 				& $0.841 \pm 0.056$		& Yes\\ 
CD-24 13837		& B1V			& B89	& 				& $0.858 \pm 0.046$		& Yes\\ 
HD 164933		& B1/2 I/II			& H88	&				& $0.810 \pm 0.066$		& Yes\\ 
\hline
\end{tabular}
\end{table*}

\subsection{Combining membership information}
\label{s-combine}

To identify a high-confidence sample of young stars in NGC~6530 and the surrounding Lagoon Nebula we combine the membership information discussed above and use {\it Gaia} DR2 parallaxes as an additional check on membership. PMs and RVs are not used to determine membership to prevent kinematic bias. After cross-matching all the data we have 4080 candidate members, with the following information:

\begin{itemize}
\item 1525 sources only have spectroscopic information from GES or the literature, or photometric H$\alpha$ emission and so their membership is determined as outlined in Section~\ref{s-spec}, giving 214 members. 
\item 1860 sources lack this data but have membership information from \citet{broo13}, which we use as given (see Section~\ref{s-nonspec}), providing 1270 members. 
\item 557 sources have both spectroscopic and \citet{broo13} membership information. In general these two sources show very good agreement, with the only difference being 25 sources identified as members by \citet{broo13} that have GES spectroscopy that suggests their surface gravities are more consistent with them being background giants (based on the $\gamma$ quantity, see Section~\ref{s-spec}), which we therefore list as non-members\footnote{This also implies that $\sim$5\% of the stars classified by \citet{broo13} as members could be non-member background giants. Our parallax cut (below) will have removed the majority of these from our sample but it is worth noting for future studies that use the \citet{broo13} membership lists.}. From the 557 sources we find 523 to be members based on the combined information.
\item 138 sources have neither spectroscopy nor membership probabilities from \citet{broo13}, and so are listed as non-members as they are currently unclassified (these are GES sources without spectroscopic measurements useful for membership determination).
\end{itemize}

\begin{figure}
\begin{center}
\includegraphics[height=240pt, angle=270]{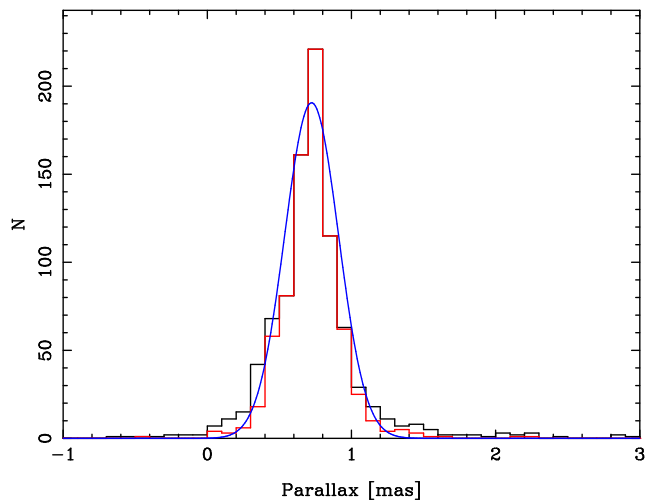}
\caption{Parallax distribution for 896 candidate members of NGC~6530 with parallaxes, selected from GES and MYStIX candidate source lists (black histogram). The blue line shows a Gaussian fit to the distribution with centre $0.724 \pm 0.006$~mas and standard deviation $0.186 \pm 0.024$~mas. The red histogram shows the distribution of 761 members that pass our parallax cut (cut levels not shown as they depend on the individual parallax errors), as described in the text.}
\label{parallax_distribution}
\end{center}
\end{figure}

This process left us with 2014 candidate members. As a last step we implemented a parallax cut using data from {\it Gaia} DR2. Of the 2014 candidate members, 896 have astrometry from {\it Gaia}, as shown in Figure~\ref{parallax_distribution}, with a clear peak at $0.724 \pm 0.006$~mas. We estimate the additional systematic error to be 0.041~mas based on the spatial distribution of our sources and the table of spatial covariances provided for {\it Gaia} DR2. This corresponds to a distance of $1326^{+77}_{-69}$~pc (accounting for the -0.03~mas zero-point offset in the {\it Gaia} parallaxes). This is in good agreement with most previous estimates \citep[e.g.,][]{pris05}, though it is significantly larger than the {\it Hipparcos}-based estimate of 600~pc from \citet[][though this was based on only 7 stars]{lokt01} and smaller than the photometric estimate of 1.8~kpc by \citet{sung00}. The parallax distribution is reasonably modelled by a Gaussian with a standard deviation of $0.186 \pm 0.024$~mas. Accounting for and subtracting the contribution of the nonuniform parallax errors using the method of \citet{ivez14}, this leaves a parallax dispersion 0.084~mas. At a distance of 1326~pc this is equivalent to $\sim$125~pc, though this is unlikely to represent the true physical dispersion, but instead may be due to under-estimated or spatially-correlated uncertainties. Based on this we exclude as members all stars with parallaxes that have $| \varpi - 0.724 | > 2 \sqrt{\sigma_\varpi^2 + 0.084^2}$, where $\varpi$ and $\sigma_\varpi$ are the measured parallax and its uncertainty. A cut at 2$\sigma$ was chosen to balance the need to include the majority of members whilst rejecting a reasonable number of non-members. This cut excluded 135 sources from our membership list, leaving 761 members with parallaxes (shown in Figure~\ref{parallax_distribution}) and 1118 members without parallaxes (either too faint to be detected by {\it Gaia} or with poor astrometry).

Our final act is to remove all stars that lack both RVs and PMs \citep[predominantly faint near-IR sources from][]{broo13}, leaving 889 sources for our kinematic study.

\subsection{Final sample of members}
\label{s-finalmembers}

\begin{figure}
\begin{center}
\includegraphics[height=240pt, angle=270]{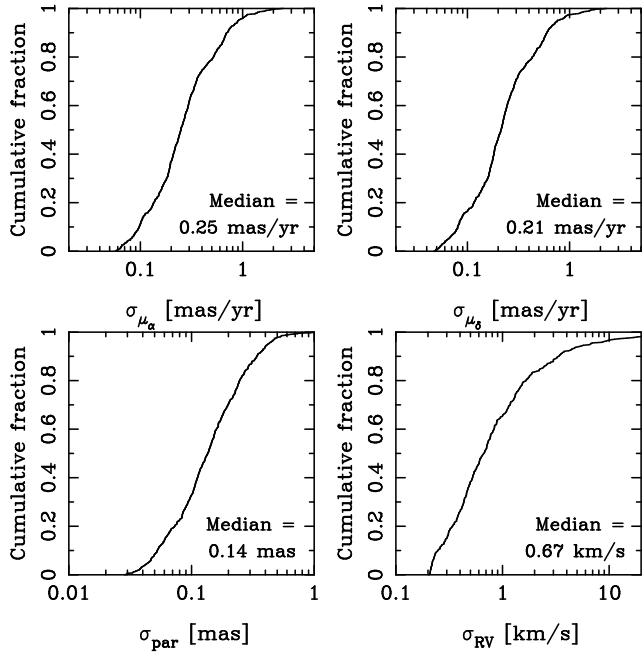}
\caption{Cumulative uncertainty distributions for our sample of members for PMs (761 stars), parallax (761 stars) and RVs (404 stars).}
\label{uncertainties}
\end{center}
\end{figure}

Our final sample of 889 members are shown in Figure~\ref{cmd} in a $G$ vs $G-J$ colour-magnitude diagram, which shows that the sample falls broadly between the 0.1 and 10~Myr pre-main sequence stellar isochrones, as found by previous studies \citep[e.g.,][]{pris07}, and as such there is no need to perform additional photometric filtering of the data. We calculated individual stellar masses for the stars in our sample by comparing the available photometry to \citet{mari17} stellar isochrones, assuming a distance of 1326~pc and an extinction of $A_V = 1.1$~mag \citep{sung00}\footnote{Note than \citet{pris19} find that most cluster members share a similar reddening so it is not unreasonable to assume a single extinction value for estimating stellar masses.}. The mass distribution of sources in our sample increases from $\sim$20~M$_\odot$ to $\sim$0.7~M$_\odot$ where it appears to turn over, with the lowest-mass stars in our sample having masses of 0.1--0.2~M$_\odot$. The majority of stars in our sample have masses of 0.4--1.0~M$_\odot$.

Of the 889 members, 404 have RVs from GES or the literature and 761 have PMs from {\it Gaia} DR2, with 276 sources having both RVs and PMs providing 3D kinematics. Figure~\ref{uncertainties} shows the cumulative uncertainty distributions for RV, PM and parallax. The median RV uncertainty is 0.67~km~s$^{-1}$, with some measurements with a precision as low as 0.25~km~s$^{-1}$. The median PM uncertainties are 0.25 and 0.21 mas~yr$^{-1}$ in RA and Dec (after adjusting for the under-estimated PM uncertainties), respectively, which at a distance of 1326~pc are equivalent to 1.5 and 1.3~km~s$^{-1}$.

Our full catalogue of members, including their photometry, astrometry, and spectroscopic parameters is included in an online table made available on Vizier.

\section{Kinematic analysis}

\begin{figure*}
\begin{center}
\includegraphics[width=460pt]{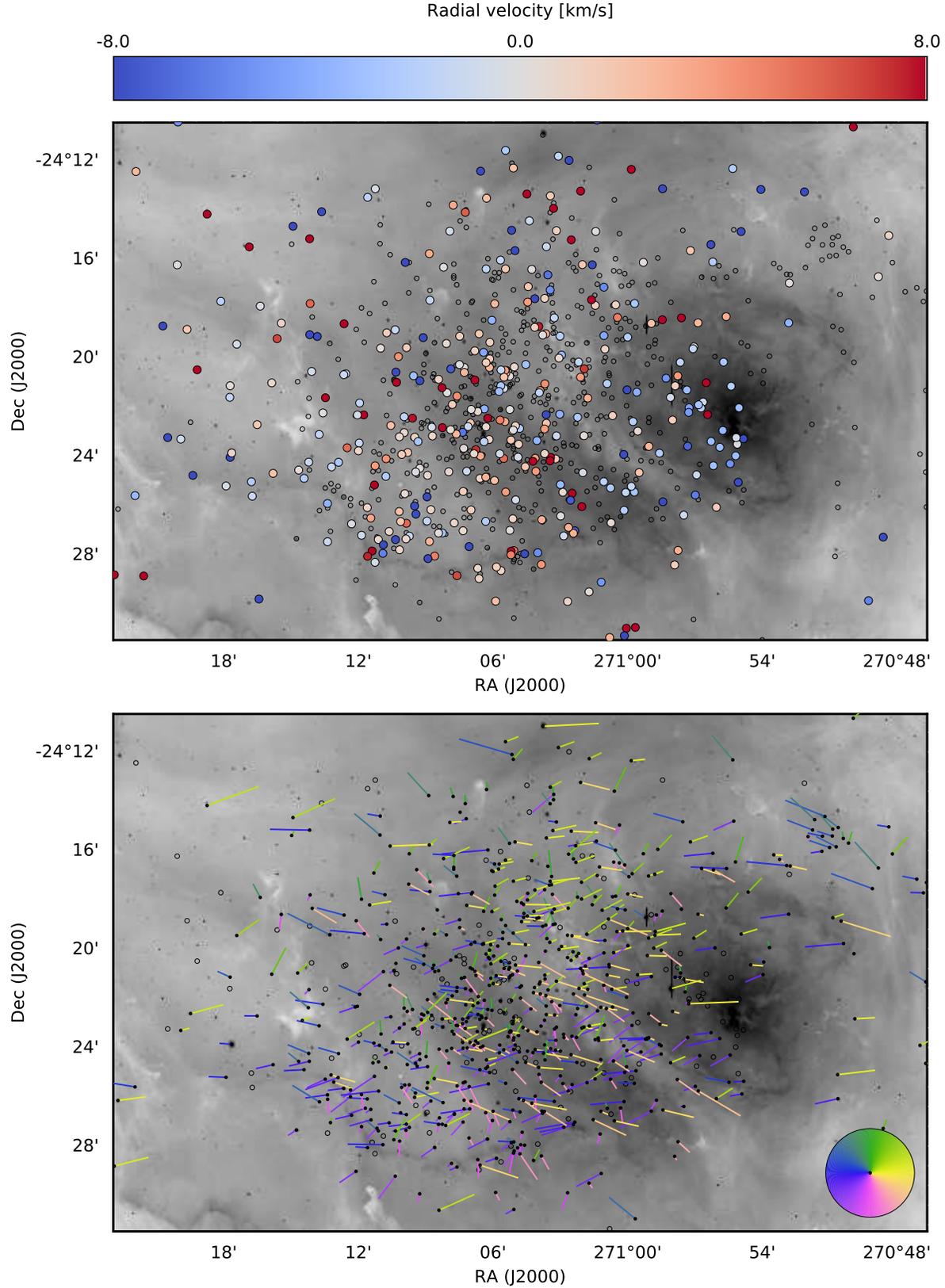}
\caption{Kinematics of our sample of 889 young stars across the Lagoon Nebula projected onto an inverted VPHAS+ H$\alpha$ image. {\it Top:} Spatial distribution of 404 sources colour-coded by their RV in the range $-8 < RV / \mathrm{km \, s}^{-1} < 8$ (sources outside this range are coloured by the limiting values of the colour bar). The 485 sources without RVs are shown as empty circles. {\it Bottom:} Spatial distribution of 761 sources with vectors showing their PMs over 0.05~Myr, relative to the median PM of $(\mu_\alpha, \mu_\delta) = (1.27, -1.98)$~mas/yr. The PM vectors are colour-coded by their PM position angle, as indicated by the colour wheel in the lower right corner. The 128 sources without usable PMs are shown as empty circles.}
\label{all_kinematics}
\end{center}
\end{figure*}

Here we use the assembled kinematic data to study the dynamics of young stars in NGC~6530 and across the surrounding Lagoon Nebula. Figure~\ref{all_kinematics} gives an overview of the kinematics of young stars across the region. It is immediately clear that while both the RVs and PMs appear generally random within the RV range and PM reference frame shown, there is some evidence for substructure. This is evident on both small scales, where small and localised groups of stars have similar velocities to each other (for example the north-western group highlighted in Figure~\ref{map_lagoon}), and on larger scales where hints of velocity gradients are evident particularly in the PM vector map. There is also evidence from Figure~\ref{all_kinematics} that the PMs have a preference for east-west motion, as evidenced by their colours. At first glance this might suggest evidence for expansion of the population in this direction, though as we will show later the expansion is primarily in the north-south direction (Section~\ref{s-velocitygradients}) and that the pattern seen here is primarily due to the larger velocity dispersion in the east-west direction (Section~\ref{s-velocityfits}).

\subsection{Kinematic outliers}
\label{s-kinematicoutliers}

A number of stars in the Lagoon Nebula have velocities that are significantly different from the general population, which we refer to as {\it kinematic outliers}. These stars may have recently been ejected from the region due to dynamical interactions or they may be contaminating non-members, but regardless, they usually do not probe the kinematics of the main population that we are interested in.

To identify these objects we calculate simple estimates of the central velocity and dispersion in each dimension using the median velocity and inter-quartile range (IQR), which is related to the standard deviation of a normal distribution as $\sigma = 0.741 \, \mathrm{IQR}$. When the velocity distributions do not follow normal distributions the IQR provides a useful, outlier-resistant method for estimating the velocity dispersion of the bulk of a population. From this we calculate central (median) velocities of ${\rm RV}_0 = 0.37$ km~s$^{-1}$ and $(\mu_\alpha, \mu_\delta)_0 = (1.27, -1.98)$ mas~yr$^{-1}$ and approximate velocity dispersions\footnote{The velocity dispersion $\sigma_{\mu_\delta}$ calculated here is smaller than that derived from a full model in Section~\ref{s-velocityfits}, despite the latter accounting for the effects of uncertainties. This is because the IQR provides only a simplistic proxy to the dispersion, but this is sufficient to identify kinematic outliers.} of $\sigma_{\rm RV} = 2.89$ km~s$^{-1}$ and $(\sigma_{\mu_\alpha}, \sigma_{\mu_\delta}) = (0.705, 0.446)$ mas~yr$^{-1}$. Stars that fall more than 3$\sigma$ from the median velocity are identified as kinematic outliers\footnote{Note that when this PM cut is transferred to physical velocity space it will include a dependence on distance and causing it to vary if there is a large spread in distance. However, the resulting change in velocity thresholds are smaller than velocity uncertainties derived from the PMs, and thus not significant.}. This gives 95 RV outliers and 66 PM outliers (35 and 47 in the two PM directions, and 16 being outliers in both PM directions), significantly more than would be expected from a normal distribution.

Since single-epoch RVs are heavily affected by binarity, the RV outliers may be dominated by true members of the Lagoon Nebula population that are in close binary systems. PMs are not affected by binarity to such a degree and therefore these outliers are expected to either not be part of the Lagoon Nebula population or to have recently been ejected from the region in some sort of dynamical interaction. The spatial distribution of the PM outliers shows no evidence for being clustered like the majority of our sources are, while their distribution in the colour-magnitude diagram (Figure~\ref{cmd}) shows a slight preference for occupying the blue-edge of the distribution, suggesting that they might not be members of the Lagoon Nebula population. Based on this, and since both non-members and recently-ejected sources will not help us resolve the dynamical state of the Lagoon Nebula, we have not included the 66 PM outliers in the kinematic analysis that follows, but have retained the RV outliers (this may introduce a bias to the RV distribution, but this is likely to be a very small effect).

\subsection{Expansion and rotation}
\label{s-expansionrotation}

\begin{figure}
\begin{center}
\includegraphics[height=230pt, angle=270]{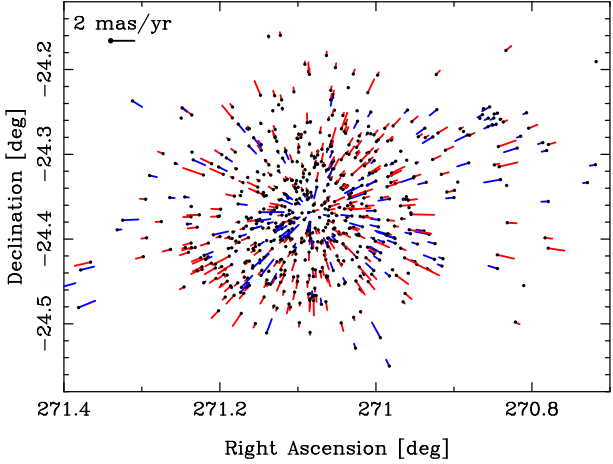}
\caption{PM vector map showing the radial component of the PMs relative to the centre (median position) of our sample. The dots show the current positions of the stars while the vectors show the PM over 0.05~Myr colour-coded red if the stars are moving outwards and blue if they are moving inwards.}
\label{expansion}
\end{center}
\end{figure}

To study the large-scale dynamics of the Lagoon Nebula population we first search for evidence of expansion, contraction and rotation of the entire population using the method of \citet{wrig16}. To do this we separate the PMs of our sources into their radial and transverse components using the median positions of stars in the population, $(\alpha, \delta) = (271.080, -24.369)$~deg. Using the stellar masses calculated in Section~\ref{s-finalmembers} we find that the kinetic energy in the PMs is split approximately equally between the radial and transverse directions with the former constituting $51 \pm 2$ (statistical) $\pm \, 3$ (positional) \% of the energy\footnote{The statistical uncertainty is calculated from a Monte Carlo experiment varying the masses and PMs of all sources by their uncertainties, while the positional uncertainty is calculated by varying the centre used by $\pm$0.1~degrees in each dimension.}.

The transverse component of the PMs shows a preference for rotation with $65.2 \pm 2.4$ (stat.) $\pm \, 4.3$ (pos.) \% of the kinetic energy in the clockwise direction. The radial component of the PMs show a significant preference for expansion however, with $76.3 \pm 1.7$ (stat.) $\pm \, 6.7$ (pos.) \% of the kinetic energy in the form of expansion. This is in agreement with \citet{kuhn19} who found that the PM velocity vectors in their sample were primarily orientated away from the cluster centre. Figure~\ref{expansion} shows the radial component of the PM vectors colour-coded as to whether they are moving towards (in blue) or away (in red) from the centre of the region. The preference for expansion is just discernible in this figure as a preference for vectors coloured red rather than blue (note that this figure highlights vector length or proper motion rather than kinetic energy).

\subsection{Velocity gradients}
\label{s-velocitygradients}

\begin{figure}
\begin{center}
\includegraphics[height=230pt, angle=270]{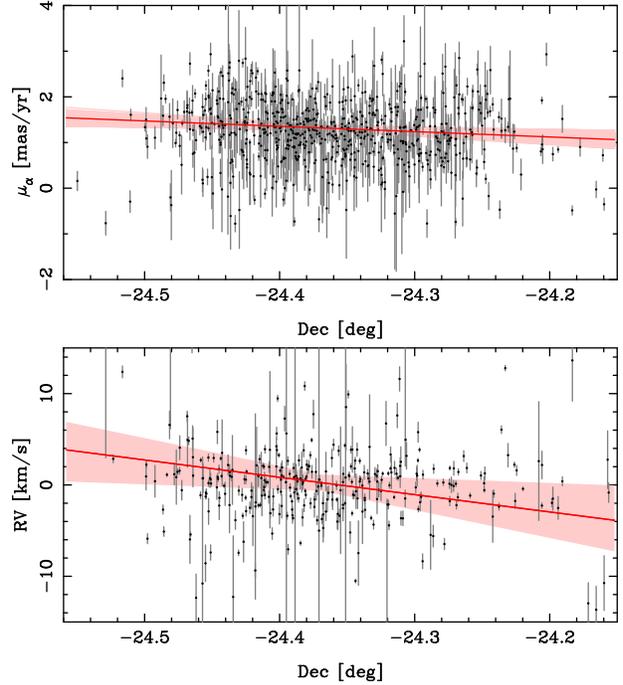}
\caption{Velocity gradients that probe significant levels of rotation. The top panel shows $\mu_\alpha$ plotted against declination and the bottom panel shows RV plotted against declination (error bars also shown). The red lines show the best-fitting velocity gradients of $-1.17 \pm 0.33$ mas~yr$^{-1}$~deg$^{-1}$ ($\mu_\alpha$) and $-19.0 \pm 12.3$ km~s$^{-1}$~deg.$^{-1}$ (RV), with the light red bands show the range of slopes within 1$\sigma$ of the best-fitting value.}
\label{rotation_gradients}
\end{center}
\end{figure}

To further explore the possible expansion of the Lagoon Nebula population we searched for velocity gradients in our kinematic data. Velocity gradients have been observed in multiple star forming regions \citep[e.g.,][]{rigl16,sacc17}, as well as in dense gas tracers \citep{andr07}, and can be attributed to processes such as rotation or expansion. We searched for velocity gradients by fitting linear relationships of the form $v = A x + B$ between the velocity, $v$, and spatial position, $x$, in each combination of dimensions. The gradient, $A$, and zero-point, $B$, were fitted by maximising the likelihood function, using the MCMC ensemble sampler {\it emcee} to sampler the posterior distribution. A third parameter ($f$) was introduced to represent the scatter in the relationship \citep[see][]{hogg10} and which was marginalised over to calculate the fit and uncertainties on the fitted velocity gradients.

\begin{figure}
\begin{center}
\includegraphics[height=230pt, angle=270]{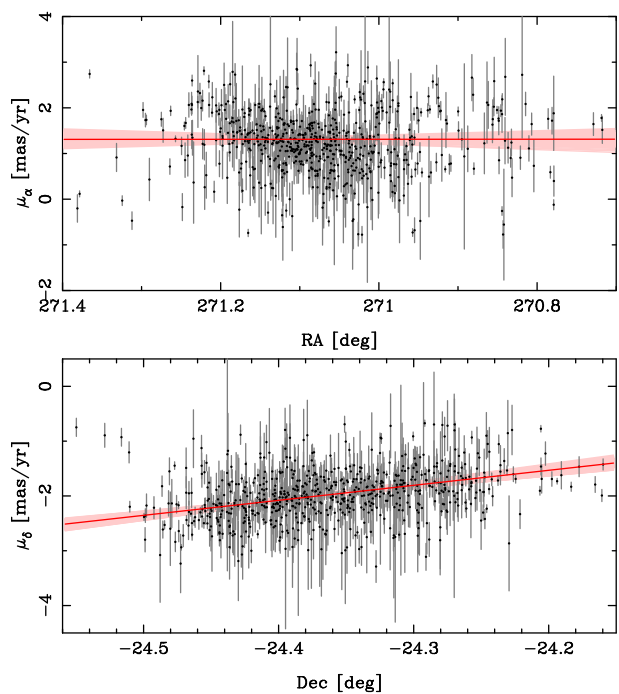}
\caption{Velocity gradients that probe expansion and contraction. The top panel shows the PM in the right ascension direction plotted against the right ascension and the bottom panel shows the PM in the declination direction plotted against the declination (error bars also shown). The red lines show the best-fitting velocity gradients of $0.0 \pm 0.21$ and $2.74 \pm 0.21$ mas~yr$^{-1}$~deg.$^{-1}$ in right ascensions and declination respectively, with the light red bands showing the range of slopes within 1$\sigma$ of the best-fitting value.}
\label{gradients}
\end{center}
\end{figure}

\begin{table}
\caption{Velocity gradient fit results for all three kinematic axes as a function of RA and Declination (the line of sight spatial distribution within the region is not resolved so we do not fit against distance). The conversion to physical units was calculated using our parallax-derived distance of 1326~pc.}
\label{gradient_fit_results} 
\begin{tabular}{lccl}
\hline
Velocity				& versus RA					& versus Dec			& Units	\\
\hline
$\mu_\alpha$			& $0.00 \pm 0.21$				& $-1.17 \pm 0.33$		& mas~yr$^{-1}$~deg$^{-1}$\\
					& $0.00 \pm 0.057$				& $-0.317 \pm 0.090$	& km~s$^{-1}$~pc$^{-1}$\\
\hline
$\mu_\delta$			& $-0.02 \pm 0.14$				& $2.74 \pm 0.21$		& mas~yr$^{-1}$~deg$^{-1}$\\
					& $-0.004 \pm 0.038$			& $0.744 \pm 0.059$		& km~s$^{-1}$~pc$^{-1}$\\
\hline
RV					& $0.92 \pm 8.3$				& $-19.0 \pm 12.3$	 	& km~s$^{-1}$~deg$^{-1}$\\
					& $0.041 \pm 0.37$				& $-0.82 \pm 0.55$		& km~s$^{-1}$~pc$^{-1}$\\
\hline
\end{tabular}
\end{table}

The results of the velocity gradient fits are listed in Table~\ref{gradient_fit_results} and can be divided into those that reveal expansion or contraction (a velocity gradient where the velocity is correlated with the position in the same dimension) and those that highlight some degree of rotation or residual angular momentum (where a velocity in one dimension is correlated with position in another dimension). Figure~\ref{rotation_gradients} shows that there is evidence for some degree of rotation or residual angular momentum within the population, particularly in $\mu_\alpha$ as a function of declination ($>$3$\sigma$ significance), with weak evidence for a correlation between RV and declination ($\sim$1.5$\sigma$ significance). The correlation between $\mu_\alpha$ and declination is consistent with the hint of clockwise rotation found in Section~\ref{s-expansionrotation}.

Figure~\ref{gradients} shows the evidence for velocity gradients indicative of expansion or contraction. Notably there is very strong evidence of a clear correlation between $\mu_\delta$ and declination with a slope of $2.74 \pm 0.21$ mas~yr$^{-1}$~deg$^{-1}$ (significance $>$10$\sigma$), but no evidence for a correlation between $\mu_\alpha$ and RA, for which we fit a slope of $0.0 \pm 0.21$ mas~yr$^{-1}$~deg$^{-1}$. The velocity gradient in declination has a positive slope, meaning that stars on the northern (southern) side of the nebula have slightly higher (lower) PMs in declination, which implies expansion, in agreement with the results of Section~\ref{s-expansionrotation}. The lack of a velocity gradient in RA doesn't change if we limit ourselves to a sample of stars with $astrometric\_excess\_noise = 0$, for which we obtain a fit of $0.64 \pm 0.75$ mas~yr$^{-1}$~deg$^{-1}$.

Our results show an interesting disagreement with those presented by \citet{kuhn19}. Those authors show a correlation between the median plane-of-sky expansion velocity and the radius from the cluster centre, with a slope of $0.6 \pm 0.2$ km~s$^{-1}$~pc$^{-1}$. They argue that this is evidence for a radially-dependent expansion velocity showing a ``Hubble flow'' like expansion pattern. However, we find that the correlation between velocity and position only exists in declination and not in RA. \citet{kuhn19} do claim to observe correlations between both $\mu_\alpha$ and RA and between $\mu_\delta$ and declination, though they do not present fits to them and the correlation in RA is ambiguous. The differences between our work and theirs may be due to their smaller sample size, their more compact area studied, the presence of contaminants (identified from GES spectroscopy) or the different quality cuts applied to the data (including their use of cuts not recommended in {\it Gaia} data release papers or by DPAC). We have shown that the expansion pattern in the Lagoon Nebula is not fully radially-dependent and instead shows a distinct asymmetry. This argues against a simple explosive expansion pattern, which we will discuss further in Section~\ref{s-discussion}.

\subsection{Velocity distribution of the entire population}
\label{s-velocitydispersion}

\begin{figure*}
\begin{center}
\includegraphics[height=460pt, angle=270]{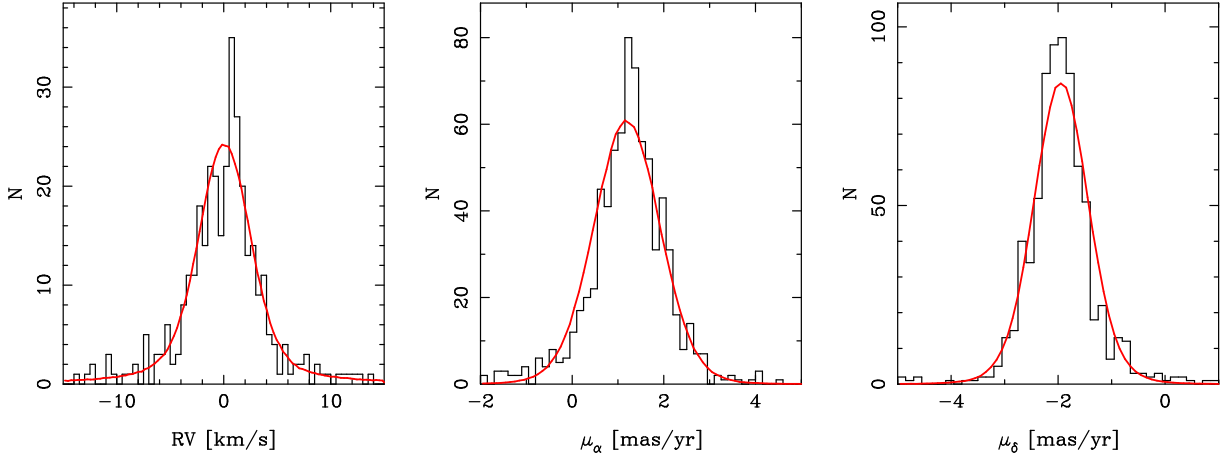}
\caption{Velocity distributions (black histograms) for our sample of young stars in RV and PMs compared to the best-fitting velocity dispersion model fit results (red distributions). The histograms show 337 out of 404 stars with RVs and 743 ($\mu_\alpha$) and 740 ($\mu_\delta$) out of 761 stars with PMs (the remainder, in all cases, fall outside of the plotted range). The model fit distributions were calculated by sampling the best-fitting forward models 1,000,000 times and plotting the resulting probability distribution functions scaled to the observations.}
\label{velocity_dispersions}
\end{center}
\end{figure*}

Figure~\ref{velocity_dispersions} shows the distribution of measured velocities in RV and PM for the stars in our sample. While at first glance the velocity distributions may appear to follow a broadly normal distribution, there is evidence for deviations from this in the form of central concentrations in velocity space in all three dimensions and hints of a double peak in the RV distribution. Shapiro-Wilk tests of normality conducted on the velocity distributions confirm this with all three dimensions rejecting the null hypothesis of a normal distribution at greater than 3$\sigma$. We suggest in Section~\ref{s-discussion} that these deviations from normality may be due to a core-halo structure with the central NGC~6530 cluster (which is both spatially and kinematically more compact, see Section~\ref{s-ngc6530}) responsible for the core of the velocity distribution and the wider Lagoon Nebula population making up the halo\footnote{A core-halo velocity distribution could also arise from the combination of a narrow, intrinsic velocity dispersion and a broad distribution of measurement uncertainties. However, the presence of kinematic subgroups within the halo distribution (Section~\ref{s-subgroups}) argues against this.}. In this section however we start by considering the velocity distribution of the entire population.

\subsubsection{Forward modelling method}
\label{s-forwardmodel}

We calculate the velocity dispersions of our sample using Bayesian inference by modelling the distribution of observed velocities using a simple parameterised model that we compare with the observations in a probabilistic way \citep[see e.g.,][]{wrig16,wrig18}. The aim of this process is to determine which of the various sets of parameters, $\boldsymbol{\theta}$, best explain the observations, $\boldsymbol{d}$. In Bayes's theorem this is known as the posterior distribution, $\boldsymbol{P(\theta | d)} = \boldsymbol{P(d | \theta)} \, \boldsymbol{P(\theta)} / \boldsymbol{P(d)}$, where $\boldsymbol{P(d | \theta)}$ is the likelihood model, $\boldsymbol{P(\theta)}$ are the priors (which includes our a priori knowledge about the model parameters) and $\boldsymbol{P(d)}$ is a normalising constant. Bayesian inference is necessary for this because it allows the model predictions to be projected into observational space, where the measurement uncertainties are defined. This is particularly important when the observations have correlated uncertainties, as is the case here. Our forward model begins by constructing a population of $N = 10^5$ stars with 3D velocities randomly sampled from a trivariate Gaussian velocity distribution along the radial and two transverse directions (the large size of the modelled population reduces the noise in the modelled velocity distributions).

Unresolved binary systems can broaden the observed RV distribution due to the contribution that binary orbital motion makes to the measured velocity. To simulate this process we follow \citet{oden02} and \citet{cott12} by assuming that a fraction of our sample are in binary systems (we don't consider triple systems because their properties are poorly constrained and are typically hierarchical, meaning that the third star is usually on a wide, long-period orbit that does not introduce a large velocity offset). The fraction of binary stars in young star clusters is poorly constrained and so we set the binary fraction to be 46\%, appropriate for solar-type field stars \citep{ragh10}. The primary star masses were sampled from the observed stellar masses, while the secondary masses were selected from a power-law companion mass ratio distribution with index $\gamma = -0.5$ over the range of mass ratios $q = 0.1$--$1.0$ \citep{regg11}. The orbital periods were selected from a log-normal distribution with a mean period of 5.03 and a dispersion of 2.28 in log$_{10}$ days \citep{ragh10}. The eccentricities were selected from a flat distribution from $e = 0$ to a maximum that scales with the orbital period as proposed by \citet{park09}. We then calculate instantaneous velocity offsets for the primary and secondary stars (relative to the centre of mass of the system) at a random phase within the binary's orbit, and then use the luminosity-weighted average of the two as the photo-centre RV\footnote{This process compensates for the fact that for high mass-ratio binaries some of the light from the secondary will contribute to the spectral features used to measure the RV, and thus the measured RV will be intermediate between that of the two stars. From simulations we found that this reduces the broadening of the RV distribution due to binaries by $\sim$25\%.}, which is then added to the modelled RV.

Finally we add measurement uncertainties for the RVs and PMs for each star, randomly sampling these from the observed uncertainty distributions and include the correlated PM uncertainties quoted in {\it Gaia} DR2 (random sampling of uncertainties is necessary because the modelled population is significantly larger than the observed population). This model has 6 free parameters, the velocity dispersion and central velocity in each dimension. Wide and uniform priors were used for each of these parameters covering $0$ to $100$ km~s$^{-1}$ for the velocity dispersions and $-100$ to $+100$ km~s$^{-1}$ for the central velocities.

To sample the posterior distribution function we use the affine-invariant Markov-Chain Monte Carlo (MCMC) ensemble sampler \citep{good13} {\it emcee} \citep{fore13} and compared the model to the observations using an unbinned maximum likelihood test, which is made efficient by the smooth velocity distributions modelled. For the MCMC sampler we used 1000 walkers and 2000 iterations, discarding the first half as a burn-in. The six parameters were found to have similar autocorrelation lengths, with the longest being $\sigma_{\mu_\alpha}$ with a length of 104 iterations, resulting in $\sim$20 independent samples per walker. The posterior distribution functions follow a normal distribution, and thus the median value was used as the best fit, with the 16$^{th}$ and 84$^{th}$ percentiles used for the 1$\sigma$ uncertainties.

\subsubsection{Results}
\label{s-velocityfits}

The results are illustrated in Figure~\ref{velocity_dispersions}. The best-fitting velocity dispersions are $\sigma_{\rm RV} = 2.12^{+0.24}_{-0.22}$ km~s$^{-1}$ and $(\sigma_{\mu_\alpha}, \sigma_{\mu_\delta}) = (0.644^{+0.044}_{-0.045}, 0.443^{+0.036}_{-0.038})$~mas~yr$^{-1}$ (full fit details in Table~\ref{velocity_fit_results}), which at a distance of 1326~pc equate to $4.06^{+0.37}_{-0.36}$ and $2.79^{+0.29}_{-0.28}$ km~s$^{-1}$ (uncertainties take into account full distance uncertainties). The three velocity distributions are significantly different from each other, particularly the RV and $\mu_\alpha$ dispersions, which imply anisotropy with a confidence of 2.3$\sigma$ (based on the difference between the RV and $\mu_\alpha$ dispersions). The binary fraction of young stars is very poorly constrained and thus the RV dispersion could be under-estimated if we have over-estimated the binary fraction. However, this effect is not large enough to increase the RV dispersion to that of the PMs -- setting the binary fraction to 0\% results in a RV dispersion of $2.47^{+0.29}_{-0.24}$~km~s$^{-1}$, still significantly smaller than the $\mu_\alpha$ dispersion.

\begin{figure}
\begin{center}
\includegraphics[height=240pt, angle=270]{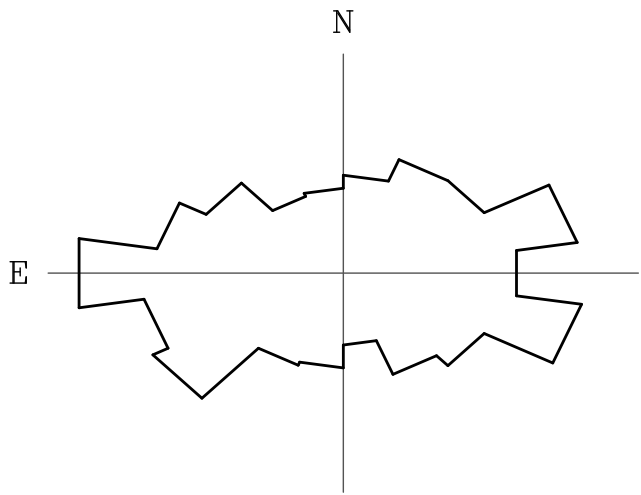}
\caption{PM position angle distribution plotted as a circular histogram with north up and east to the left.}
\label{theta}
\end{center}
\end{figure}

The PM velocity dispersions are also significantly different from each other and provide equally strong evidence for anisotropy (with a significance of 2.2$\sigma$ between them). Figure~\ref{theta} shows the distribution of PM vector position angles on the plane of the sky, which shows that the PM velocity distribution is ellipsoidal, with a semi-major axis closely aligned with the right ascension axis (the larger velocity distribution in this direction goes a long way to explaining the preference for east-west motion as shown in Figure~\ref{all_kinematics}). The velocity anisotropy implies the region is not sufficiently dynamically mixed to have developed an isotropic velocity dispersion. Such anisotropy has been observed in other regions, particularly OB associations, which are believed to be dynamically un-evolved \citep{wrig16,wrig18}. In OB associations the anisotropy can often be attributed to kinematic substructure within the region, which may be the case in the Lagoon Nebula (see Section~\ref{s-subgroups}). 

Our PM velocity dispersions are larger than those reported by \citet{kuhn19} of $2.7 \pm 0.2$ and $1.8 \pm 0.2$ km~s$^{-1}$. This is because their sample covers a smaller area on the sky than ours, being limited to the area of the central {\it Chandra} observations included in the MYStIX survey, and so is dominated by the core of the velocity distribution that originates from the central NGC~6530 cluster (see Section~\ref{s-ngc6530}). Our PM velocity dispersions are significantly smaller than the $0.85 \pm 0.06$ mas~yr$^{-1}$ velocity dispersion measured by \citet{chen07b}\footnote{The authors only quote a 3D velocity dispersion of $1.48 \pm 0.14$ mas~yr$^{-1}$, from which we derive an estimate of their measured 1D velocity dispersion.} from photographic plates. This could either be because the two samples have different membership or because the photographic plates may suffer from distortions. 

Our combined three-dimensional velocity dispersion was calculated as $\sigma_{3D} = \sqrt{\sigma_{\rm RV}^2 + \sigma_{\mu_\alpha}^2 + \sigma_{\mu_\delta}^2} = 5.35^{+0.39}_{-0.34}$~km~s$^{-1}$.

\begin{table*}
\caption{Central velocities and velocity dispersions for the entire Lagoon Nebula population and the subsets of NGC~6530, M8E and the north-western subgroup. Velocity dispersions are not calculated for the latter two as they are not sufficiently well sampled. The north-western subgroup lacks any sources with spectroscopy and so its RV is not known.}
\label{velocity_fit_results} 
\begin{tabular}{lcccccc}
\hline
Sample								& RV$_0$				& $\sigma_{\rm RV}$		& $\mu_{\alpha_0}$		& $\sigma_{\mu_\alpha}$	& $\mu_{\delta_0}$		& $\sigma_{\mu_\delta}$	\\
									& [km~s$^{-1}$]		& [km~s$^{-1}$]		& [mas~yr$^{-1}$]		& [mas~yr$^{-1}$]		& [mas~yr$^{-1}$]		& [mas~yr$^{-1}$]		\\
\hline
Lagoon Nebula (all members)				& $0.08^{+0.34}_{-0.30}$	& $2.12^{+0.24}_{-0.22}$	& $1.19 \pm 0.05$		& $0.644^{+0.044}_{-0.045}$	& $-1.94 \pm 0.04$	& $0.443^{+0.036}_{-0.038}$		\\
NGC 6530 ($r < 2.5^\prime$)				& $0.82^{+0.34}_{-0.34}$	& $1.80^{+0.40}_{-0.37}$	& $1.21 \pm 0.04$		& $0.251^{+0.045}_{-0.043}$	& $-2.00 \pm 0.03$	& $0.216^{+0.038}_{-0.036}$\\
M8E ($r < 2^\prime$)					& $0.75 \pm 0.72$		& -					& $1.78 \pm 0.23$		& -					& $-2.14 \pm 0.34$	& -					\\
North-western group ($r < 1^\prime$)		& -					& -					& $1.77 \pm 0.20$		& -					& $-1.50 \pm 0.17$	& -					\\
\hline
\end{tabular}
\end{table*}

\subsubsection{Virial state}
\label{s-virialstate}

To measure the virial state of the Lagoon Nebula stellar population we use the virial equation, which in its three-dimensional form is given by

\begin{equation}
\sigma_{3D}^2 = \frac{GM_{vir}}{2 r_{vir}} ,
\end{equation}

\noindent where $\sigma_{3D}$ is the three-dimensional velocity dispersion, $G$ is the gravitational constant, $M_{vir}$ is the virial mass and $r_{vir}$ is the virial radius. We substitute the parameter $\eta = 6 r_{vir} / r_{eff}$, where $r_{eff}$ is the effective (or half-light) radius, and rearrange to give the virial mass as

\begin{equation}
M_{vir} = \eta \frac{\sigma_{3D}^2 \, r_{eff} }{3G}
\end{equation}

\noindent where the parameters $\eta$ and $r_{eff}$ are determined by fitting an \citet[][hereafter EFF]{elso87} surface brightness profile to the stellar distribution. We do this with our sample and find parameters of $\eta = 9.3 \pm 0.1$ and $r_{eff} = 4.83^{+0.73}_{-0.48}$ arcmin, which equates to $1.87^{+0.41}_{-0.30}$~pc at a distance of 1326~pc.

This gives a virial mass of $3.8^{+1.0}_{-0.7} \times 10^4$ M$_\odot$. The total stellar mass in the Lagoon Nebula has been estimated to be between 1000 \citep{pris05,toth08} and 4000~M$_\odot$ \citep{kuhn15b}. Based on our full sample of likely members, which we estimate to be complete in the range of 0.7--3.0~M$_\odot$, and extrapolating out using a \citet{masc13} initial mass function, we estimate the total mass to be $\sim$2500~M$_\odot$. The mass of the Lagoon Nebula H{\sc ii} region and parental molecular cloud in the immediate vicinity of the stellar population are not well constrained, though the mass of the molecular clouds in the wider area have been estimated to be 2--6 $\times 10^4$~M$_\odot$ \citep{take10}. Such a mass of material would be sufficient to keep the stellar system in virial equilibrium, but since the stellar population is not embedded within these molecular clouds (as evidenced by the relatively low extinction towards most members) this material will not contribute to the virial state of the system. We therefore conclude that the stellar population within the Lagoon Nebula is gravitationally unbound and should continue to expand in the future.

\subsection{Velocity distribution of subgroups within the Lagoon Nebula}
\label{s-subgroups}

\begin{figure*}
\begin{center}
\includegraphics[height=460pt, angle=270]{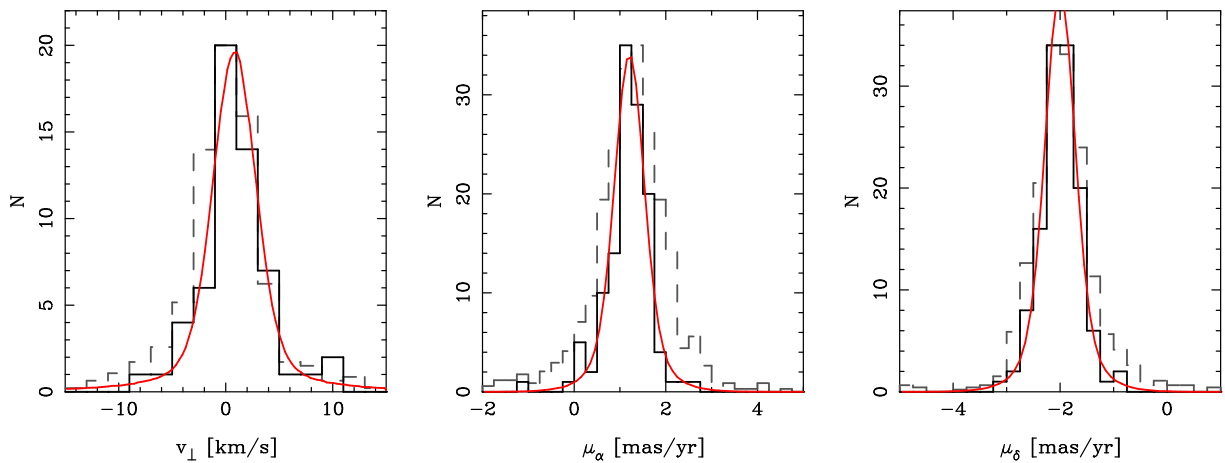}
\caption{Velocity distributions (black histograms) for a sample of 150 young stars in the NGC~6530 cluster (within a radius of 150$^{\prime\prime}$) in RV and PMs. The histograms show 57 out of 60 stars with RVs (the remainder fall outside the plotted range) and all 126 stars with PMs. In red we show the best-fitting velocity dispersion model fits for the NGC~6530 cluster population. The model fit distributions were calculated by sampling the best-fitting forward model 1,000,000 times and plotting the resulting probability distribution functions scaled to the observations. The velocity ranges plotted are the same as those in Figure~\ref{velocity_dispersions} and we also show for comparison the velocity distributions for all stars across the Lagoon Nebula (grey dashed histograms) as shown in that figure.}
\label{subset_dispersions}
\end{center}
\end{figure*}

The velocity dispersion calculations above consider stars across the Lagoon Nebula, including the central NGC~6530 cluster, a number of known subgroups, and a wider less-clustered population (see Figure~\ref{map_lagoon}). If these substructures represent real clusters they may be kinematically distinct from the rest of the Lagoon Nebula population and could be gravitationally bound, even if the wider population is not. We therefore considered the kinematics of four groups of stars within the Lagoon Nebula complex, three that are commonly known (NGC~6530, M8E, and the Hourglass Nebula cluster) and one that is clear from the spatial distribution of X-ray sources and was first identified by \citet{kuhn14}. We selected stars in each of these groups based on the regions shown in Figure~\ref{map_lagoon}, which were themselves based on the central positions identified by \citet[][with the exception of NGC~6530, see below]{kuhn14}. For each group we then use a 2-sample Anderson-Darling (AD) test and the tables of \citet{scho87} to verify the null hypothesis that the stars in each group have kinematics drawn from the same distribution as the wider population.

\subsubsection{M8E, the Hourglass Nebula and the north-western subgroup}

For the M8E subgroup we selected 38 stars within a radius of 2$^\prime$. These stars have similar RVs and $\mu_\delta$ velocities to the wider population, but have a preference for more eastward PMs relative to stars across the rest of the nebula (as evident from Figure~\ref{all_kinematics}). AD tests confirm this, providing a p-value of 0.0004 for $\mu_\alpha$ and allowing us to reject the null hypothesis. The M8E subgroup therefore appears kinematically distinct.

For the Hourglass Nebula we selected 6 stars within a radius of 1$^\prime$, but found that their kinematics were similar to that of the main population. The AD tests do not provide significant evidence that the velocities of stars are not drawn from that of the wider distribution. Since the Hourglass Nebula cluster is highly embedded, our optically-selected sample probably only includes sources projected against the cluster and therefore may not be representative of the Hourglass Nebula population.

We identify 14 stars as likely members of the north-western subgroup, all within a radius of 1$^\prime$. The majority of these stars have PMs in the north-easterly direction (as seen in Figure~\ref{all_kinematics}) that marks them out as kinematically distinct (though sparsely sampled). AD tests provide support for this, with p-values of 0.0031 ($\mu_\alpha$), and 0.0023 ($\mu_\delta$) implying that the null hypothesis can be rejected and that the group is kinematically distinct.

The central (median) velocities of these subgroups are listed in Table~\ref{velocity_fit_results} (with the exception of the Hourglass Nebula population, which we believe has not been properly sampled), though we do not calculate velocity dispersions as the groups are too sparsely sampled.

\subsubsection{The central NGC 6530 cluster and its virial state}
\label{s-ngc6530}

The NGC~6530 cluster sits at the centre of the Lagoon Nebula and we selected a sample of 150 stars in a region with a radius of 150$^{\prime\prime}$ centred on (RA, Dec) = (271.105, -24.371)~deg. The kinematics of these stars exhibit a narrower velocity distribution than that of the entire population (see Figure~\ref{subset_dispersions}), particularly in $\mu_\alpha$, and are offset towards slightly larger RVs. AD tests provide strong evidence for this with p-values of 0.036 (RV), 0.00084 ($\mu_\alpha$), and 0.019 ($\mu_\delta$) that allow us to reject the null hypothesis that the kinematics of these stars are drawn from the same distribution as the wider population. Shapiro-Wilk tests of normality conducted on the velocity distributions provide no evidence that the PM distributions are inconsistent with following normal distributions, while the RV distribution does deviate significantly from normality, most likely due to the presence of binaries. The three spatial subclusters within the NGC~6530 cluster that were identified by \citet{kuhn14} appear to be kinematically indistinct based on our data, with no evidence from the AD tests that they are not drawn from the same velocity distributions as each other, so we don't subdivide our sample any further. We fit the velocity dispersion of this subset using the same method as for the entire sample.

The results are shown in Figure~\ref{subset_dispersions} and listed in Table~\ref{velocity_fit_results}. The best-fitting velocity dispersions are $\sigma_{\rm RV} = 1.80^{+0.40}_{-0.37}$~km~s$^{-1}$ and $(\sigma_{\mu_\alpha}, \sigma_{\mu_\delta}) = (0.251^{+0.045}_{-0.043}, 0.216^{+0.038}_{-0.036})$~mas~yr$^{-1}$, which at a distance of 1326~pc equate to $1.58^{+0.31}_{-0.29}$ and $1.36^{+0.26}_{-0.24}$ km~s$^{-1}$ (uncertainties take into account the distance uncertainties). Contrary to the whole sample these three velocity dispersions are in agreement with each other within the uncertainties (significance of anisotropy $< 1 \sigma$), suggesting an isotropic velocity dispersion that implies the cluster has been sufficiently mixed to erase any primordial anisotropy. The 3D velocity dispersion is $2.75^{+0.39}_{-0.28}$~km~s$^{-1}$.

We fit an EFF profile in the same manner described earlier and find an effective radius of $r_{eff} = 1.83^{+0.42}_{-0.35}$ arcminutes or $0.71^{+0.20}_{-0.17}$~pc at a distance of 1326~pc, and $\eta = 9.3 \pm 0.1$. Combined with the 3D velocity dispersion this equates to a virial mass of $3800^{+1100}_{-900}$~M$_\odot$. Extrapolating the stellar population to a full initial mass function we estimate the stellar mass of the central NGC~6530 cluster to be $480 \pm 100$~M$_\odot$, which is an order of magnitude lower than the virial mass. Therefore, despite the central cluster having an isotropic velocity dispersion that would imply it is dynamically mixed, the cluster is not in virial equilibirum and should disperse in the future.

\section{Discussion}
\label{s-discussion}

Here we discuss our results and their implications for understanding young star clusters. Our results can briefly be summarised as follows.

\begin{itemize}
\item There is strong evidence for expansion of the Lagoon Nebula stellar population, with 76\% of the kinetic energy in the radial component of the PMs in the direction of expansion. However this expansion is preferentially in the Declination direction, as evidenced by the strong velocity gradient between $\mu_\delta$ and Declination, and there is no evidence for a systematic expansion pattern in the right ascension direction.
\item The 3D velocity distributions are non-Gaussian, with a strong central peak due to the NGC~6530 cluster, and the velocity dispersions are anisotropic, with the velocity dispersion in the right ascension direction being significantly larger than that in the declination direction. The 3D velocity dispersion is $5.35^{+0.39}_{-0.34}$~km~s$^{-1}$, giving a virial mass of $\sim$38,000~M$_\odot$, an order of magnitude larger than the known stellar mass, implying that the region is globally unbound.
\item The central cluster NGC~6530 has a 3D velocity dispersion of $2.75^{+0.39}_{-0.28}$~km~s$^{-1}$, which gives a virial mass of $\sim$3,800~M$_\odot$, also about an order of magnitude larger than the known stellar mass and implying the cluster is unbound. The velocity dispersion is consistent with isotropy and the PM distributions follow a normal distribution, suggesting the cluster is dynamically mixed.
\end{itemize}

The broad picture that the kinematics give us is that the stellar population is anisotropic, gravitationally unbound and expanding, though this expansion is not isotropic and (at least within the plane of the sky) is preferentially occurring in the declination direction (though since the region is gravitationally unbound it should expand in the future in both directions).  We discuss the two main results here.

\subsection{Current and past dynamical states}

The non-normal velocity distributions show a central peak that appears to be due to the kinematically-distinct NGC~6530 cluster in the centre of the Lagoon Nebula, and potentially other kinematic substructures that are less-well sampled. This core-halo velocity pattern, where the core cluster exhibits a narrower velocity distribution than the wider halo population, has been observed in other young regions \citep[e.g.,][]{jeff14}. In this scenario the denser NGC~6530 cluster would have formed within a larger, lower-density and unbound region that we now see as the Lagoon Nebula stellar population. The dispersal of the residual gas left over from star formation may have affected the dynamics of both groups of stars, particularly if either was gravitationally bound.

The velocity dispersions of both the entire stellar population and the central NGC~6530 cluster are sufficiently large that they appear to be gravitationally unbound at the moment. It is not clear whether the entire region or the NGC~6530 cluster was ever gravitationally bound in the past. Gravitationally bound stellar groups tend to be well mixed, as the necessary high stellar density for virial equilibrium leads to a short dynamical timescale. The central NGC~6530 cluster does have an isotropic and Gaussian velocity dispersions that suggests it is well mixed and therefore may have been a bound cluster in the past.

For the wider Lagoon Nebula population the non-Gaussian velocity distributions and clear anisotropy suggest that the system has not been sufficiently mixed and therefore was not a bound cluster in the past. The semi-major axis of the anisotropic velocity distribution on the plane of the sky is orientated almost east-west, roughly aligned with the spatial elongation of the nebula itself, and, to a lesser extent, the stellar system. This may be an indication that the velocity distribution and anisotropy are primordial and have not been well mixed.

Some N-body simulations predict that velocity anisotropy can arise in star clusters following violent relaxation in a tidal field \citep{vesp14}. However the main signal of this anisotropy is predicted to occur at radii of between 2--5 half-mass radii, significantly beyond the area covered by our current sample (which extends to approximately 2 and 3 half-mass radii in declination and right ascension respectively), and that the inner regions should be well mixed after such events. These simulations also model the initial distribution of stars as simple spherical and centrally concentrated systems, whereas young star forming regions are considerably more substructured \citep[e.g.,][]{cart04,schm11} and as such further work that considers more realistic initial conditions may be required.

\subsection{Expansion of the Lagoon Nebula population}

The cause of the observed expansion across the Lagoon Nebula is unclear. The most commonly cited mechanism for the disruption and expansion of a bound cluster is residual gas expulsion, whereby feedback from young stars disperses the gas left over from star formation leaving the remaining stellar part of the system in a super-virial state and prone to dissolution \citep[e.g.,][]{tutu78,lada84}. The stars in the Lagoon Nebula are very young, but are no longer embedded in their parental molecular cloud and \citet{dami17} identify a shell of ionized gas expanding towards us away from the region that suggests feedback and therefore residual gas expulsion could be a possible cause of expansion. However, early work on residual gas expulsion predicted a symmetric and radial expansion of the unbound cluster, which is inconsistent with the asymmetric expansion observed here. More recent theoretical studies actual suggest that gas expulsion doesn't always lead to the stars becoming unbound \citep[e.g.,][]{krui12b,park13}, which also argues against this being the cause of the observed expansion pattern.

Preferential expansion in one direction could be caused by tidal disruption of the cluster by a nearby giant molecular cloud, an idea which has been explored in detail by \citet{elme10} and \citet{krui11b}. \citet{take10} mapped out the molecular clouds in the vicinity of the Lagoon Nebula, identifying massive (each $\sim$2--6 $\times 10^4$~M$_\odot$) clouds centred on the Hourglass Nebula and M8E, as well as clouds to the south-west and north of the main stellar population. The location of molecular clouds approximately to the north and south of the Lagoon Nebula population might therefore support the role of tidal heating in driving the expansion of the stellar population.

An alternative explanation for the asymmetric expansion pattern is that it is a relic of the formation process (either of the stars or the cluster itself) and that the stellar population has not been sufficiently mixed to erase it. If the region formed from the collision between two molecular clouds in such a way that the remnant cluster was gravitationally unbound and retained a kinematic imprint of the formation process then this could lead to an apparent expansion pattern along the collision axis. Another possibility is that the region formed from the merger of multiple sub-clusters along a certain axis and is now expanding along that axis following violent relaxation \citep[e.g.,][]{park16}.

The main difference between these scenarios and those of asymmetric residual gas expulsion or tidal heating is whether the expansion pattern is due to a cluster-unbinding mechanism that can produce this asymmetry or whether the kinematics are a remnant of an asymmetry that was present before the star and cluster formation process began (and possibly played some role in the formation). The latter theory does require that the cluster has not been sufficiently-well mixed to have erased this primordial asymmetry, but the non-normal and anisotropic velocity dispersions do suggest that such mixing has not yet occurred. Distinguishing between these ideas will either require improved simulations of these process to compare to existing observations or observations of the gas-phase kinematics to search for evidence of a collision.

\section{Summary}

We have performed a 3D kinematic study of the young stellar population in the Lagoon Nebula and centred around the young cluster NGC~6530. This was performed using a combination of {\it Gaia}-ESO Survey spectroscopy that provides RVs and {\it Gaia} DR2 astrometry that provides PMs, as well as various indicators of youth used to identify the stellar population. This led to the following findings.

\begin{enumerate}
\item Using Bayesian inference and forwarding modelling we have calculated the 3D velocity dispersion of the Lagoon Nebula stellar population to be $5.35^{+0.39}_{-0.34}$~km~s$^{-1}$, which gives a virial mass more than an order of magnitude larger than the known stellar mass. This implies that the system is gravitationally unbound.
\item The velocity dispersion is anisotropic and the velocity distributions are not well represented by Gaussians, implying that the region is not fully mixed on large scales. The PM velocity dispersion in the right ascension direction is significantly larger than in the declination direction, with the PM velocity ellipsoid aligned east-west.
\item There is moderate evidence for rotation and strong evidence that the region is expanding, with 76\% of the kinetic energy in the PM radial component in the direction of expansion. However, the expansion is not symmetric and is almost entirely in the declination direction.
\item The central NGC~6530 cluster has a more compact velocity distribution with a dispersion of $2.75^{+0.39}_{-0.28}$~km~s$^{-1}$. This also implies a virial mass that is much larger than the known stellar mass, though there is evidence the cluster might have been bound in the past.
\end{enumerate}

To conclude, the stellar population in the Lagoon Nebula is gravitationally unbound and shows clear evidence for expansion. However, the expansion pattern is not consistent with simple residual gas expulsion models that predict a radial expansion pattern. This suggests either a more complex gas expulsion process, tidal disruption from nearby giant molecular clouds, or that the stellar population was never gravitationally bound and that the expansion pattern observed is a remnant of the primordial gas or stellar kinematics from before the cluster formed. The combination of a larger velocity dispersion in the right ascension direction, but a stronger expansion pattern in the declination direction is intriguing and suggests the region is not (and likely never has been) dynamically relaxed and mixed. These results add to the debate over the impact and effectiveness of residual gas expulsion on the dynamics of young star clusters and stellar systems \citep[e.g.,][]{krui12b,park13,wrig14b,wrig18,ward18}. It is clear that the simple picture of embedded star clusters in virial equilibrium being disrupted by residual gas expulsion cannot reproduce the complex observational picture coming from recent kinematic studies. It remains to be seen whether more complex simulations  of the effects of gas expulsion on embedded star clusters can match the observations, or whether all such systems start their lives from more complex initial conditions.

\section{Acknowledgments}

NJW acknowledges an STFC Ernest Rutherford Fellowship (grant number ST/M005569/1). RJP acknowledges support from the Royal Society in the form of a Dorothy Hodgkin Fellowship. AB acknowledge support from ICM (Iniciativa Cient\'ifica Milenio) via the N\'ucleo Milenio de Formaci\'on Planetaria. EJA acknowledges support from the Spanish Government Ministerio de Ciencia, Innovacion y Universidades though grant AYA2016-75 931-C2-1and from the State Agency for Research of the Spanish MCIU through the "Center of Excellence Severo Ochoa" award for the Instituto de Astrofisica de Andalucia (SEV-2017-0709). This research has made use of NASA's Astrophysics Data System and the Simbad and VizieR databases, operated at CDS, Strasbourg. The authors would like to thank Jim Dale and Alexis Klutsch for comments and discussion on this work.

This work is based on data products from observations made with ESO Telescopes at the La Silla Paranal Observatory under program ID 188.B-3002. These data products have been processed by the Cambridge Astronomy Survey Unit (CASU) at the Institute of Astronomy, University of Cambridge, and by the FLAMES-UVES reduction team at INAF-Osservatorio Astrofisico di Arcetri. These data have been obtained from the Gaia-ESO Survey Data Archive, prepared and hosted by the Wide Field Astronomy Unit, Institute for Astronomy, University of Edinburgh, which is funded by the UK Science and Technology Facilities Council. This work also made use of results from the European Space Agency (ESA) space mission Gaia. Gaia data are being processed by the Gaia DPAC. Funding for the DPAC is provided by national institutions, in particular the institutions participating in the Gaia MultiLateral Agreement (MLA). The Gaia mission website is https://www.cosmos.esa.int/gaia. The Gaia archive website is https://archives.esac.esa.int/gaia. This paper also makes use of public survey data (programme 177.D-3023, the VST Photometric H$\alpha$ Survey of the Southern Galactic Plane and Bulge) obtained at the European Southern Observatory.

\bibliographystyle{mn2e}
\bibliography{/Users/nwright/Documents/Work/tex_papers/bibliography.bib}
\bsp

\end{document}